\documentclass[%
 aip,
 amsmath,amssymb,
preprint,%
]{revtex4-1}

\usepackage{graphicx}% Include figure files
\usepackage{dcolumn}% Align table columns on decimal point
\usepackage{bm}% bold math
\usepackage[utf8]{inputenc}
\usepackage[T1]{fontenc}
\usepackage{mathptmx}
\usepackage{etoolbox}

\draft % marks overfull lines with a black rule on the right

\graphicspath{{figures/}}

\newcommand{\dso}{{D\lowercase{y}S\lowercase{c}O$_3$}}
\newcommand{\etal}{{\it et al.}}
\newcommand{\gso}{{G\lowercase{d}S\lowercase{c}O$_3$}}

\newcommand{\pto}{{P\lowercase{b}T\lowercase{i}O$_3$}}

\newcommand{\sso}{{S\lowercase{m}S\lowercase{c}O$_3$}}
\newcommand{\sro}{{S\lowercase{r}R\lowercase{u}O$_3$}}
\newcommand{\sto}{{S\lowercase{r}T\lowercase{i}O$_3$}}

\begin{document}

\title{Mapping the complex evolution of ferroelastic/ferroelectric domain patterns in epitaxially strained PbTiO$_3$ heterostructures}

\author{Lichtensteiger Céline}
\email[]{Celine.Lichtensteiger@unige.ch}
%\homepage[]{Your web page}
%\thanks{}
%\altaffiliation{}
\affiliation{Department of Quantum Matter Physics, University of Geneva, 24 Quai Ernest-Ansermet, CH-1211 Geneva 4, Switzerland}

\author{Hadjimichael Marios}
%\email[]{Your email address}
%\homepage[]{Your web page}
%\thanks{}
%\altaffiliation{}
\affiliation{Department of Quantum Matter Physics, University of Geneva, 24 Quai Ernest-Ansermet, CH-1211 Geneva 4, Switzerland}

\author{Zatterin Edoardo}
%\email[]{Your email address}
%\homepage[]{Your web page}
%\thanks{}
%\altaffiliation{}
\affiliation{ESRF, The European Synchroton, 71 Avenue des Martyrs, Grenoble 38000, France}

\author{Su Chia-Ping}
%\email[]{Your email address}
%\homepage[]{Your web page}
%\thanks{}
%\altaffiliation{}
\affiliation{Université Paris-Saclay, CNRS, Laboratoire de Physique des Solides, Orsay 91405, France}

\author{Gaponenko Iaroslav}
%\email[]{Your email address}
%\homepage[]{Your web page}
%\thanks{}
%\altaffiliation{}
\affiliation{Department of Quantum Matter Physics, University of Geneva, 24 Quai Ernest-Ansermet, CH-1211 Geneva 4, Switzerland}

\author{Tovaglieri Ludovica}
%\email[]{Your email address}
%\homepage[]{Your web page}
%\thanks{}
%\altaffiliation{}
\affiliation{Department of Quantum Matter Physics, University of Geneva, 24 Quai Ernest-Ansermet, CH-1211 Geneva 4, Switzerland}

\author{Paruch Patrycja}
%\email[]{Your email address}
%\homepage[]{Your web page}
%\thanks{}
%\altaffiliation{}
\affiliation{Department of Quantum Matter Physics, University of Geneva, 24 Quai Ernest-Ansermet, CH-1211 Geneva 4, Switzerland}

\author{Gloter Alexandre}
%\email[]{Your email address}
%\homepage[]{Your web page}
%\thanks{}
%\altaffiliation{}
\affiliation{Université Paris-Saclay, CNRS, Laboratoire de Physique des Solides, Orsay 91405, France}

\author{Triscone Jean-Marc}
%\email[]{Your email address}
%\homepage[]{Your web page}
%\thanks{}
%\altaffiliation{}
\affiliation{Department of Quantum Matter Physics, University of Geneva, 24 Quai Ernest-Ansermet, CH-1211 Geneva 4, Switzerland}

\date{\today}

%\pacs{}

\begin{abstract}

We study the complex ferroelastic/ferroelectric domain structure in the prototypical ferroelectric \pto\ epitaxially strained on (110)$_o$-oriented \dso\ substrates, using a combination of atomic force microscopy, laboratory and synchrotron x-ray diffraction and high resolution scanning transmission electron microscopy. We observe that the anisotropic strain imposed by the orthorhombic substrate creates a large asymmetry in the domain configuration, with domain walls macroscopically aligned along one of the two in-plane directions. We show that the periodicity as a function of film thickness deviates from the Kittel law. As the ferroelectric film thickness increases, we find that the domain configuration evolves from flux-closure to $a/c$-phase, with a larger scale arrangement of domains into superdomains. 

\end{abstract}

\maketitle

%\tableofcontents

\section{Introduction}

In ferroelectric thin films, the interplay between mechanical and electrostatic boundary conditions allows for the formation of a large variety of domain structures with fascinating properties. This is particularly the case in \pto , a tetragonal ferroelectric with a polarisation developing along the $c$-axis mostly due to ionic displacements. In \pto\ thin films, the orientation of the polarisation and arrangement into domain structures have been theoretically studied~\cite{Pertsev-PRL-1998,Pertsev-PRL-2000,Koukhar-PRB-2001,Li-APL-2001,Jiang-PRB-2014,Chapman-PCCP-2017}, and are described as phase diagrams with regions of different domain configurations as a function of epitaxial strain and temperature (see review by Schlom \etal ~\cite{Schlom-AnnuRevMaterRes-2007}). When the electrostatic boundary conditions are modified and the depolarisation field is introduced, the polarisation configurations in \pto\ become more complex. For example, in \pto/\-\sto\ superlattices, periodic repetitions of \pto\ and \sto\, deposited on \dso\ substrates, the ferroelectric layers display ordered arrays of polar vortices~\cite{Yadav2016}. Additionally, the signature of Bloch polarisation components was observed using resonant soft x-ray diffraction (RSXD)~\cite{Shafer2018}. A ``supercrystal'' structure of very ordered flux closure domains was stabilised using ultrafast light pulses~\cite{Stoica2019}, a configuration somewhat similar to the spontaneously ordered phase observed in \pto/\sro\ superlattices deposited on \dso ~\cite{Hadjimichael-PhD-2019,Hadjimichael-NatMat-2021}. An incomensurate spin crystal was observed in \pto\ thin films between \sro\ electrodes on \dso ~\cite{Rusu-Nature-2022}. Ferroelectric skyrmions were predicted in \pto ~\cite{Goncalves2019} and subsequently measured  in \pto/\sto\ superlattices on \sto\ substrates~\cite{Das2019}, and polar merons were observed in mixed $a_1/a_2-a/c$ phase \pto\ films under tensile epitaxial strain on a \sso\ substrate~\cite{Wang-NatureMaterials-2020}. All these observations demonstrate that the interplay between elastic and electrostatic energies creates structures which can be simultaneously controlled using electric fields and light and give rise to novel phenomena, like negative capacitance~\cite{Iniguez-NatureReviewsMaterials-2019}. The studied systems are usually complicated and are characterised by mixed phases of different domain configurations~\cite{Damodaran2017}. 

In this work, we use a combination of atomic force microscopy, laboratory and synchrotron x-ray diffraction, and high resolution scanning transmission electron microscopy to study domain structures in \pto\ thin films sandwiched between top and bottom \sro\ layers, grown on (110)$_o$-oriented \dso\ substrates. We find that the anisotropic strain imposed by the orthorhombic substrate creates a large asymmetry in the domain configuration, with domain walls macroscopically aligned along one of the two in-plane directions. We show that the periodicity as a function of film thickness deviates from the Kittel law, but agrees well with the period determined in other studies for \pto\ layers in different types of heterostructures under the same epitaxial strain. As the ferroelectric film thickness increases, the domain configuration evolves from a complex flux-closure-like pattern to the standard $a/c$-phase, characterised by a larger scale arrangement of domains. Finally, we show that above a certain critical thickness, the large structural distortions associated with the ferroelastic domains propagate through the top \sro\ layer, creating a modulated structure that extends beyond the ferroelectric layer thickness. The varying length scales of these periodic phenomena reveal the hierarchy of the different energy costs at play within the \pto\ layers.

\section{Results}

A series of samples has been grown by off-axis radiofrequency (RF) magnetron sputtering on (110)$_o$-oriented \dso\ substrates, with 55 unit cells (u.c.) thick bottom and top \sro\ electrodes, and \pto\ film thicknesses ranging from 23 up to 133 u.c. (see Section.~\ref{section:Growth} for details regarding sample growth). To understand the domain configuration in such samples, both the elastic strain and electrostatic boundary conditions must be considered.

Bulk \pto\ is ferroelectric below a critical temperature of 765 K, with a tetragonal structure and lattice parameters $a = b = 3.904$ \AA\ and $c = 4.152$ \AA\ at room temperature~\cite{Shirane-ActaCrystallographica-1956}. \dso\ is orthorhombic with room temperature $Pbnm$ space group lattice parameters~\cite{Velickov-ZKristallogr-2007} $a_o = 5.443(2)$ \AA , $b_o = 5.717(2)$ \AA\ and $c_o = 7.901(2)$ \AA , corresponding to pseudocubic lattice parameters $a_{\mathrm{pc}} = c_{\mathrm{pc}} = \frac{\sqrt{a_o^2+b_o^2}}{2} = 3.947$ \AA , $b_{\mathrm{pc}} = c_o/2 = 3.951$ \AA , $\alpha_{\mathrm{pc}}=\gamma_{\mathrm{pc}}=90^\circ$, $\beta_{\mathrm{pc}}=2 \cdot \arctan({a_o/b_o})=87.187^\circ$. For (110)$_o$-oriented \dso, the out-of-plane [001]$_{\mathrm{pc}}$ direction is equivalent to [110]$_o$, while the in-plane directions [100]$_{\mathrm{pc}}$ and [010]$_{\mathrm{pc}}$ are equivalent to [$\overline{1}$10]$_o$ and [001]$_o$ respectively~\footnote{``${pc}$'' subscript refers to the pseudocubic unit cell, while ``$o$'' is used to refer to the orthorhombic unit cell.}. In our sample series, the in-plane strain imposed by \dso\ on \pto\ films at room temperature can thus be calculated as $\frac{a_{\mathrm{pc}}-a_0}{a_{\mathrm{pc}}}=-0.25\%$ along [100]$_{\mathrm{pc}}$ and $\frac{b_{\mathrm{pc}}-a_0}{b_{\mathrm{pc}}} = -0.16\%$ along [010]$_{\mathrm{pc}}$, where $a_0=3.957$ \AA\ is the equivalent lattice parameter of \pto\ in the room-temperature cubic paraelectric phase. To accommodate this strain, \pto\ thin films on \dso\ are expected to be in the $a/c$-phase, with regions where the $c$-axis points out-of-plane ($c$-domains) as well as regions where it points in-plane ($a$-domains), giving rise to a ferroelastic $a/c$-domain configuration with 90$^\circ$ domain walls, as predicted in Ref.~\cite{Koukhar-PRB-2001} and demonstrated experimentally (see for example Ref.~\cite{Catalan-NatMat-2011,Nesterov-APL-2013,Highland-APL-2014}). 

Additionally to these ferroelastic domains, the electrostatic boundary conditions also play a role. The polarisation charges at the surface of the \pto\ layers are partially screened~\cite{Junquera-NAT-2003,Aguado-Puente-PRL-2008,Stengel-NatMat-2009,Li-APL-2017,Hadjimichael-PRM-2020} by charges in the \sro\ layers. Bulk \sro\ is a metallic transition-metal oxide and is often used as an electrode in the ferroelectric oxides community~\cite{Eom-Science-1992}. It is orthorhombic with room temperature $Pbnm$ space group lattice parameters $a_o = 5.57$ \AA , $b_o = 5.53$ \AA\ and $c_o = 7.85$ \AA ~\cite{Randall-JACS-1959}, corresponding to the pseudocubic unit cell parameters $a_{\mathrm{pc}} = c_{\mathrm{pc}} = 3.924$ \AA , $b_{\mathrm{pc}} = 3.925$ \AA , $\alpha_{\mathrm{pc}}=\gamma_{\mathrm{pc}}=90^\circ$, $\beta_{\mathrm{pc}} = 90.413^\circ$. The depolarisation field arising from the incomplete screening of the surface charges can lead the $c$-domains to split in alternating ``up'' ($c^+$) and ``down'' ($c^-$) domains with 180$^\circ$ domain walls and plays a role in the domain configuration for the thinner films. The combination of mechanical and electrostatic constraints can then result in flux-closure structures, as observed in tensile-strained \pto\ thin films~\cite{Tang-Science-2015,Li-APL-2017,Li-ActaMat-2019}.

We address the question of domain configuration and evolution as a function of the \pto\ film thickness by using different techniques. We observe the pattern visible at the surface of the \sro /\pto /\sro\ heterostructures by atomic force microscopy (Section~\ref{Section:AFM}), while x-ray diffraction measurements are used to extract the domain periodicity (Section~\ref{Section:XRD}). These measurements not only demonstrate that the evolution of the domain period with film thickness deviates from the Kittel law, but also reveal an additional larger period appearing for the thicker films. The origin of this larger period is then investigated using scanning x-ray nanodiffraction microscopy with high spatial resolution, highlighting the arrangement of $a/c$ domains into superdomains (Section~\ref{Section:nanodiffraction}). The evolution of the domain configuration with increasing film thickness from flux-closure-like to $a/c$ domains is also demonstrated by direct imaging using transmission electron microscopy, where the arrangement of the $a/c$ domains into superdomains is further confirmed (Section~\ref{Section:TEM}).

\subsection{Periodic pattern at the surface of the heterostructures observed by atomic force microscopy}
\label{Section:AFM}

\begin{figure}[!htb]
\includegraphics[width=1\linewidth]{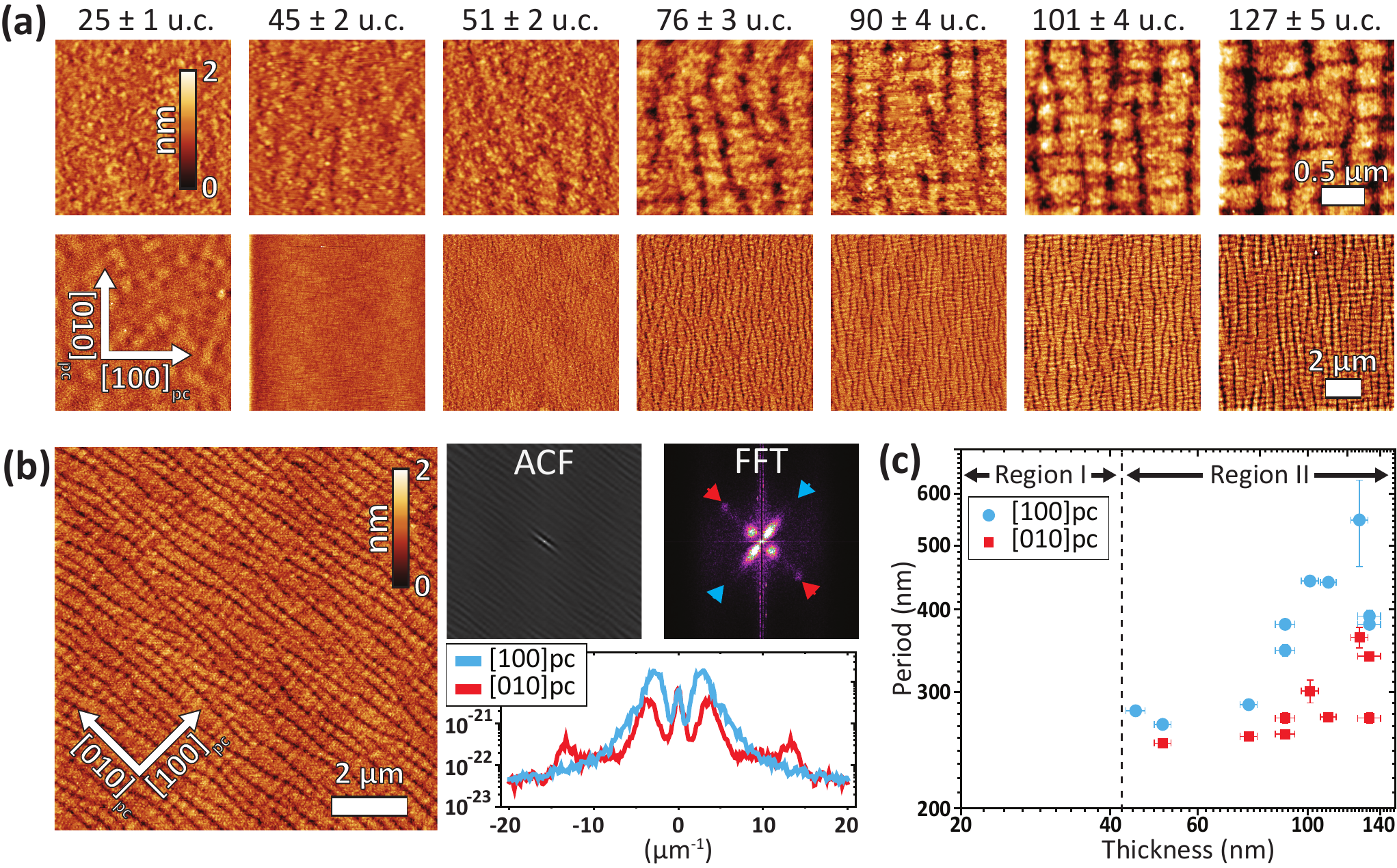}
\caption{\label{fig:Figure1} (a) AFM topography images obtained on the different samples. Top row are 2x2 $\mu$m$^2$ scans, bottom row are 10x10 $\mu$m$^2$ scans, shown in the same 0-2 nm height scale. The orientation of each sample was fixed with respect to the substrate pseudo-cubic axes [100]$_{\mathrm{pc}}$ and [010]$_{\mathrm{pc}}$. From these images, it is clear that as the \pto\ layer thickness increases, trenches develop on the surface of the \sro\ top layer, with a pattern that gets more pronounced and anisotropic with thickness, with long and deep trenches parallel to the [010]$_{\mathrm{pc}}$ axis, and smaller trenches parallel to the [100]$_{\mathrm{pc}}$ axis. (b) Extracting the period for the 90 $\pm$ 4 u.c. thick \pto\ layer using the Gwyddion software~\cite{Necas-CEJP-2012}. From the topography measurement, the autocorrelation function (ACF) is taken. A fast Fourier transform (FFT) is then applied, displaying periodic peaks along [100]$_{\mathrm{pc}}$ and [010]$_{\mathrm{pc}}$, clearly visible in the cuts, allowing us to determine the periodicity. (c) Evolution of the periodicity of the surface tilt pattern along the two crystallographic axis.}
\end{figure}

Atomic force microscopy (AFM) topography images obtained on the different samples reveal that as the \pto\ layer thickness increases, trenches develop at the surface of the \sro\ top layer in an organised pattern (Figure~\ref{fig:Figure1}). For samples of 45 u.c. and below, this pattern is hardly visible and the top \sro\ is atomically flat. The pattern gets more pronounced and anisotropic with increasing \pto\ layer thickness, with long and deep trenches parallel to the [010]$_{\mathrm{pc}}$ axis, and smaller trenches parallel to the [100]$_{\mathrm{pc}}$ axis, while the surface roughness stays reasonably low, with root mean square (RMS) roughness values ranging from 157 to 393 pm over surfaces of 10 $\mu$m $\times$ 10 $\mu$m. The pattern that we observe at the surface of the \sro\ top layer is comparable to what has been observed at the surface of \pto\ layers grown on \dso\ substrates in Ref~\cite{Nesterov-APL-2013} as a result of periodic ferroelastic $a/c$ domains.

To extract the period of the distortions visible on the surface of the samples, we calculate the autocorrelation function (ACF) of the topography image, and subsequently the fast Fourier transform (FFT) of the autocorrelation image, as shown in Figure~\ref{fig:Figure1}(b) for the sample with the 90 u.c. thick \pto\ layer. As elaborated in Ref~\cite{Nesterov-APL-2013}, this method is more sensitive to the periodic distortions of the surface of the sample than a direct FFT of the topography image. The results for the different samples are shown in Figure~\ref{fig:Figure1}(c) and also reported in the concluding Figure~\ref{fig:Figure5}. Once visible, the periods along both directions ([100]$_{\mathrm{pc}}$ and [010]$_{\mathrm{pc}}$) increase as a function of \pto\ layer thickness, with the period along [100]$_{\mathrm{pc}}$ being always larger than the period along [010]$_{\mathrm{pc}}$.

\subsection{Periodic patterns in the \pto\ layers observed by x-ray diffraction}
\label{Section:XRD}

\begin{figure}[!htb]
\includegraphics[width=\linewidth]{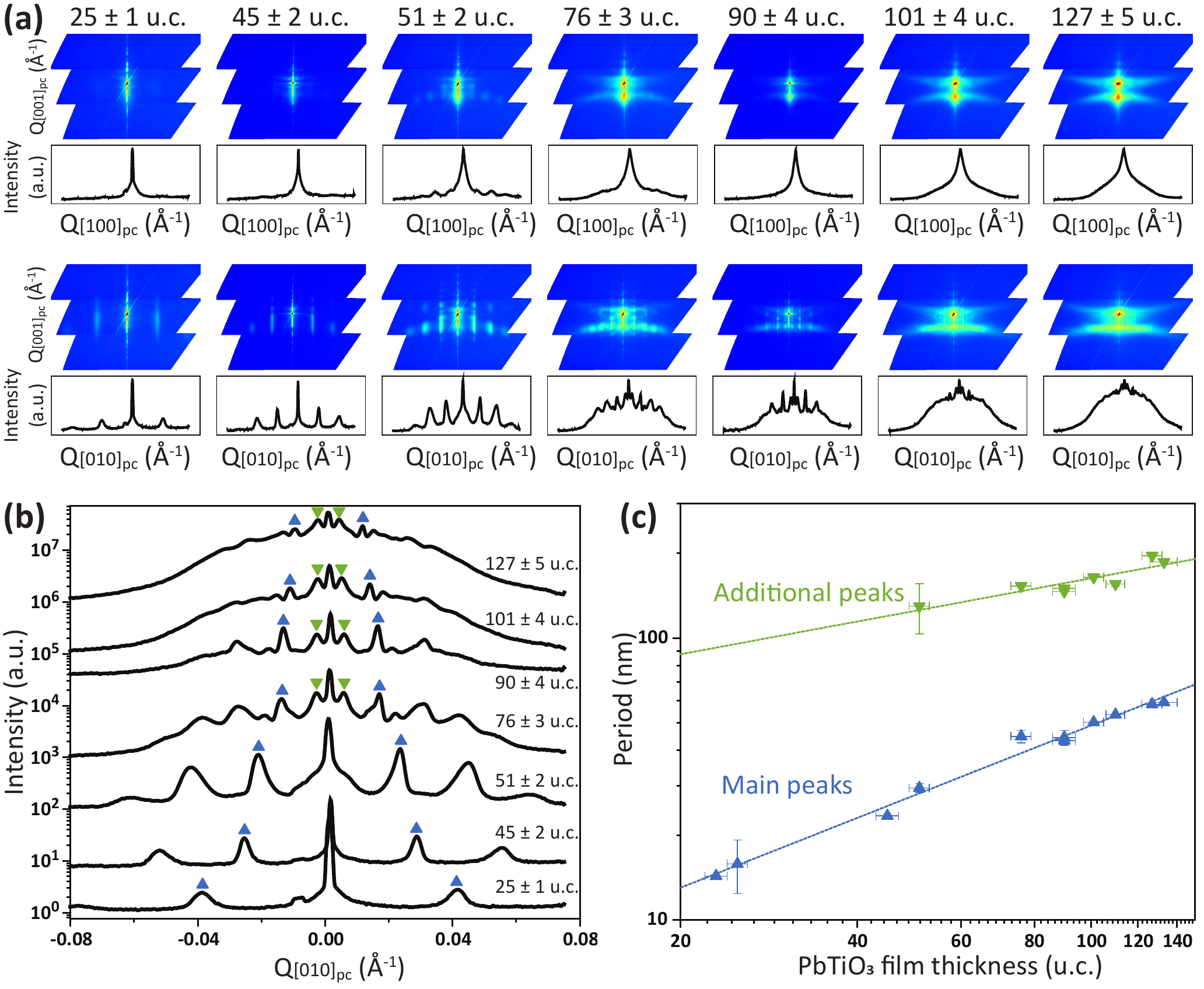}
\caption{\label{fig:XRD_RSM} (a) Reciprocal space maps (RSM) around the (001)$_{\mathrm{pc}}$ peak of the substrate in the $Q_{[001]_{\mathrm{pc}}}-Q_{[100]_{\mathrm{pc}}}$ plane (top row) and in the $Q_{[001]_{\mathrm{pc}}}-Q_{[010]_{\mathrm{pc}}}$ plane (bottom row), for samples with different \pto\ thicknesses. $Q_{[100]_{\mathrm{pc}}}$ and $Q_{[010]_{\mathrm{pc}}}$ vary from -0.1 to 0.1 \AA$^{-1}$, while $Q_{[001]_{\mathrm{pc}}}$ varies between 1.4 and 1.8 \AA$^{-1}$. Below each map is the corresponding intensity obtained from a cut at $Q_{[001]_{\mathrm{pc}}}=1.55$ \AA$^{-1}$ ($c$-domains), displaying the periodic peaks originating from the domain pattern. These intensity cuts are reported in (b) for comparison, and the periods are plotted in (c) - see discussion in the text.}
\end{figure}

To relate this pattern observed at the surface of the \sro\ layer to the domain pattern in the \pto\ layers below, we used in-house x-ray diffraction to measure reciprocal space maps (RSM) of our different samples (Figure~\ref{fig:XRD_RSM}). For the thicker films, the butterfly shape - a signature of the $a/c$-phase - can be recognised, with the high intensity peak of the substrate overlapping with the peak of the $a$-domains and the peak of \sro . The peak of the $c$ domains appears at a lower $Q_{[001]_{\mathrm{pc}}}$ value. The butterfly wings arise from the tilts in the $a$- and $c$-domains~\cite{Lee-AnnuRevMaterRes-2006}.

Comparing in Figure~\ref{fig:XRD_RSM} the RSM obtained in the $Q_{[100]_{\mathrm{pc}}}$-$Q_{[001]_{\mathrm{pc}}}$ plane (top row) and in the $Q_{[010]_{\mathrm{pc}}}$-$Q_{[001]_{\mathrm{pc}}}$ plane (bottom row), we observe again an anisotropy between the two in-plane crystallographic axes $[100]_{\mathrm{pc}}$ and $[010]_{\mathrm{pc}}$: periodic peaks are clearly visible along $[010]_{\mathrm{pc}}$, whereas the peaks along $[100]_{\mathrm{pc}}$ are less well-defined and exhibit lower intensity. The position of these peaks was determined from intensity cuts at $Q_{[001]_{\mathrm{pc}}}=1.55$ \AA$^{-1}$, corresponding to the region of the $c$-domains, reported in Figure~\ref{fig:XRD_RSM}(b) (see blue arrows). From the position of these peaks, we extracted the periodicity - plotted in blue in Figure~\ref{fig:XRD_RSM}(c) as a function of the \pto\ film thickness. The blue line serves as a guide to the eye and is obtained by fitting the evolution of the period $p$ with film thickness $t$ using an exponential expression, $p=a\cdot t^n$. The best fit was obtained with $n =0.8\pm 0.1$, deviating from the Kittel law~\cite{Landau-book-1992,Kittel-PR-1946,Mitsui-PR-1953,Roitburd-PSSA-1976,Pompe-JAP-1993,Pertsev-JAP-1995} where $n$ should be equal to $1/2$ (see Section~\ref{Section:Discussion} for a discussion regarding this deviation from the Kittel law).

Upon further analysis of the periodic peaks appearing in the RSM, we note in Figure~\ref{fig:XRD_RSM}(b) that additional peaks appear for the thicker films (see down-facing green triangles). These additional peaks correspond to larger periods, as reported in Figure~\ref{fig:XRD_RSM}(c) in green. This time, the fit gives $n=0.4\pm 0.2$, much closer to what one would expect from the Kittel law. For comparison, we analysed a 90 u.c \pto\ sample without the top \sro\ electrode using piezo-response force microscopy measurements. We observed that in addition to the $a/c$ domains, the out-of-plane polarisation arranges into larger regions alternating between up and down polarisation, forming $c^+/c^-$ superdomains (see discussion in Supporting Information Section~\ref{SI:topo_PFM_XRD_uncapped} and in Ref.~\cite{Tovaglieri-ToBeSubmitted-2023}). The period of these superdomains matches with the value obtained from the additional peak in the RSM, confirming that the additional peaks come from the arrangement of the out-of-plane polarisation in $c^+/c^-$ superdomains. These additional peaks appear only for samples with a \pto\ layer thicker than $\sim$ 45 u.c., which is also the thickness above which the pattern starts to appear in the topography of the surface layer, as observed by AFM.

%\newpage

\subsection{Tilts in $a$- and $c$-domains observed by synchrotron x-ray nanodiffraction}
\label{Section:nanodiffraction}

\begin{figure}[!htb]
\includegraphics[width=\linewidth]{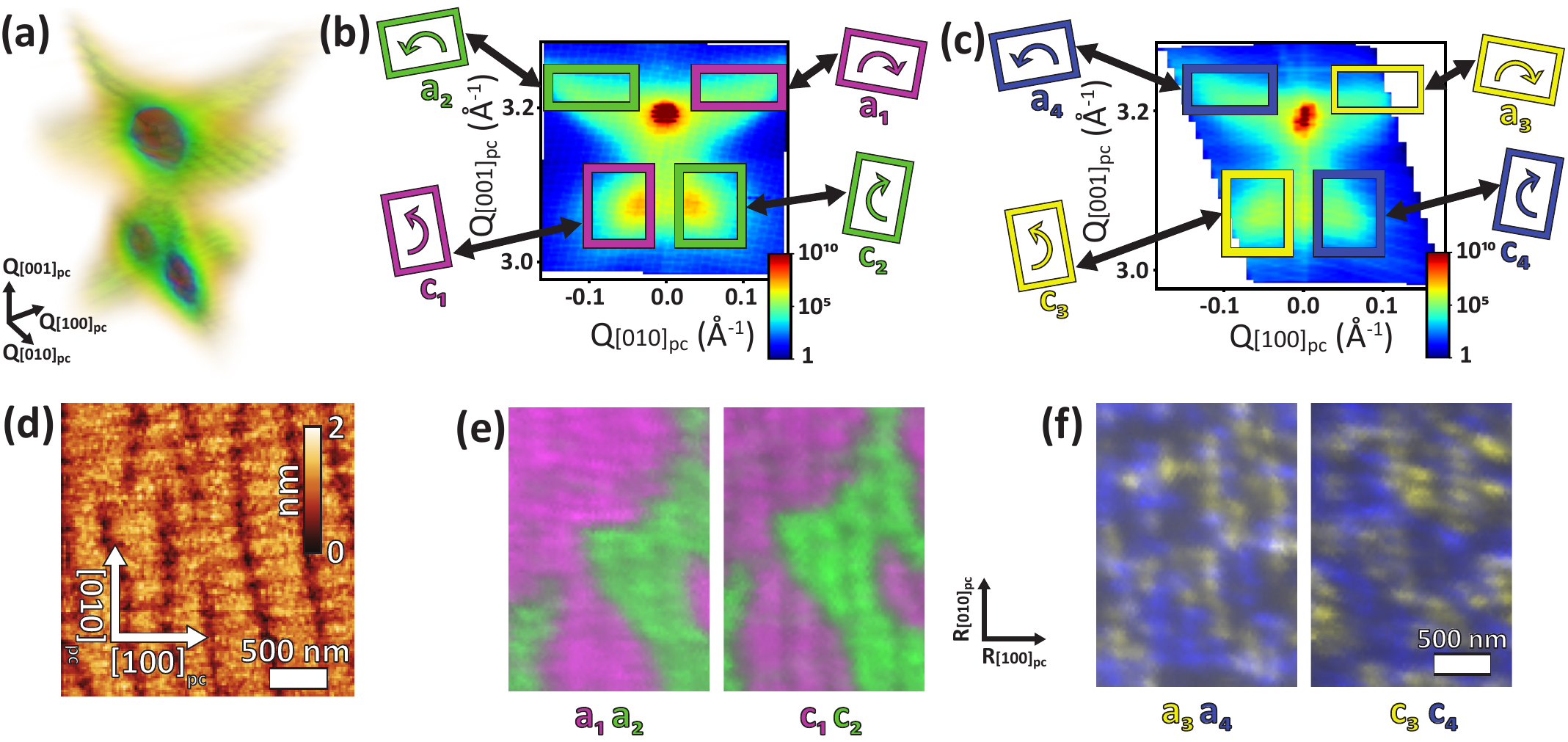}
\caption{\label{fig:Figure3} Scanning x-ray diffraction microscopy of the $002$ peak of a 90$\pm$4 u.c. thick \pto\ between top and bottom \sro\ electrodes (53$\pm$2 u.c. thick) on a \dso\ substrate. (a) Average $002$ 3D-RSM calculated by summing all local 3D-RSMs collected at different sample positions within the SXDM raster scan. (b-c) Projection along $Q_{[100]_{\mathrm{pc}}}$ (b) and $Q_{[010]_{\mathrm{pc}}}$ (c) of the average $002$ 3D-RSM. Coloured boxes define ROIs corresponding to $a$ and $c$ domains tilted in the four symmetrically equivalent $\langle100\rangle_{\mathrm{pc}}$ directions. (d) Topography of this sample displayed in the same scale used in the spatial intensity maps for comparison. (e-f) Spatial maps of the sum of the intensity scattered within the ROIs defined in (b-c) using the corresponding color scale: in (e), the intensities corresponding to $a_1$ (pink) and $a_2$ (green) are overlaid using complementary colors; the same is done for $c_1$ (pink) and $c_2$ (green); in (f), $a_3$ (yellow) and $a_4$ (blue) are overlaid; as for $c_3$ (yellow) and $c_4$ (blue). Note that complementary colors are used, so that white (respectively black) corresponds to the superposition (respectively absence) of the two domains.}
\end{figure}

The 90 u.c. thick \pto\ layer sample was also analysed using scanning x-ray nanodiffraction microscopy (SXDM). This technique allows us to measure raster scans of local 3D-RSMs, with a spatial resolution determined by the focused x-ray beam size~\cite{Hadjimichael-PRL-2018}, here $\sim 30\times30\,\mathrm{nm}^2$ full width at half maximum (FWHM). The output of an SXDM map is thus the diffracted intensity $I$ as a function of three reciprocal space coordinates ($Q_{[100]_{\mathrm{pc}}}$, $Q_{[010]_{\mathrm{pc}}}$, $Q_{[001]_{\mathrm{pc}}}$), and two direct space coordinates ($R_{[100]_{\mathrm{pc}}}$, and $R_{[010]_{\mathrm{pc}}}$) (Figure~\ref{fig:Figure3}).

 The average $002$ 3D-RSM obtained by summing $I$ over all sample positions $(R_{[100]_{\mathrm{pc}}}, R_{[010]_{\mathrm{pc}}})$ is shown in Figure~\ref{fig:Figure3}(a). The projections along $Q_{[100]_{\mathrm{pc}}}$ and $Q_{[010]_{\mathrm{pc}}}$ are shown in Figure~\ref{fig:Figure3}(b-c). This RSM displays the ``butterfly'' shape characteristic of the $a/c$ phase already evident in Figure~\ref{fig:XRD_RSM}. Compared to Figure~\ref{fig:XRD_RSM} however, in Figure~\ref{fig:Figure3}(a-c) no satellites are present: because the domain period is comparable to the nanofocused beam size in a local 3D-RSM, the number of coherently illuminated periods is insufficient to give rise to the constructive interference that generates the satellites. The coloured boxes in Figure~\ref{fig:Figure3}(b-c) each define a region of interest (ROI) corresponding to an $a$ or $c$ domain tilted in one of the four $\langle100\rangle_{\mathrm{pc}}$ directions.

The sum of the intensity scattered within each ROI is computed for each local RSM and plotted as a function of $(R_{[100]_{\mathrm{pc}}}, R_{[010]_{\mathrm{pc}}})$ in Figure~\ref{fig:Figure3}(e-f). Here, each plot is labelled with the domain type and colour of the respective ROI, allowing us to map the presence or absence of the different domains. Note that due to their small size, it is clusters, or ``bundles'' of domains rather than individual ones that are visible. By comparing the intensity maps to the topography of the sample displayed at the same scale in Figure~\ref{fig:Figure3}(d), one sees that the contrast is comparable, showing that the pattern observed in the topography of the \sro\ top layer is related to the arrangement of the $a/c$ domains in the \pto\ layer below.

Comparing the $a$ and $c$ domain maps in Figure~\ref{fig:Figure3}(e-f), we see a clear correlation between the spatial distribution of different tilts: the spatial distribution of $a_1$ domains matches that of the $c_1$ domains (pink), the $a_2$ of the $c_2$ (green), the $a_3$ of the $c_3$ (yellow) and the $a_4$ of the $c_4$ (blue). Such pairing is to be expected given the crystallography of $a/c$ twins~\cite{Speck1995}. More interestingly, $a_1 / c_1$ and $a_2 / c_2$ pairs appear to further aggregate at a larger scale into homogeneous areas, forming superdomains compatible with the observations in Figure~\ref{fig:XRD_RSM} and Figure~\ref{fig:SI_topo_PFM_XRD_uncapped}. 

\subsection{Domain patterns in \pto\ layers observed by transmission electron microscopy}
\label{Section:TEM}

\begin{figure}[!htb]
\includegraphics[width=1\linewidth]{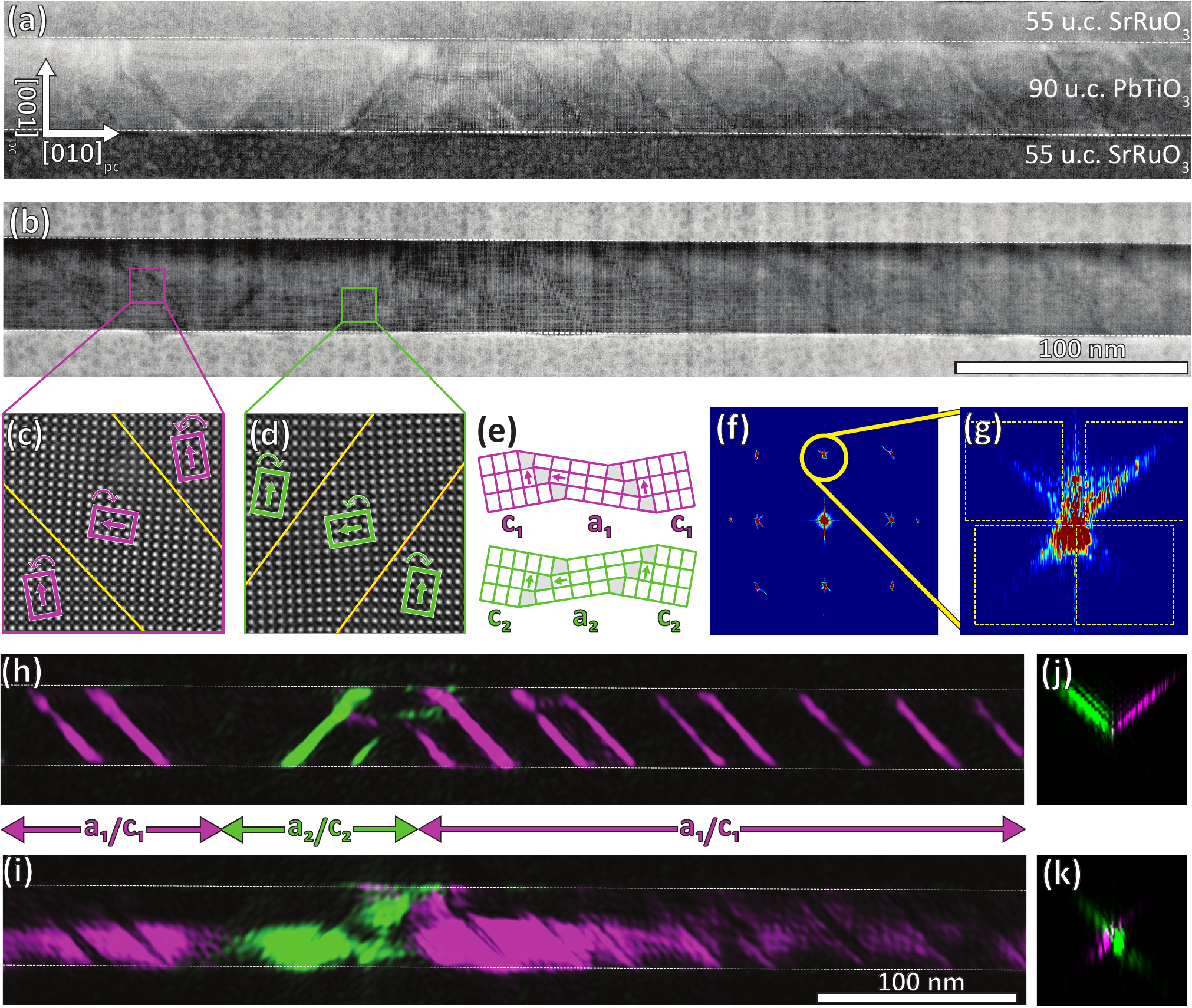}
\caption{\label{fig:Figure4} STEM image of a 90$\pm$4 u.c. thick \pto\ between top and bottom \sro\ electrodes (53$\pm$2 u.c. thick) on \dso\ substrate. (a-b) Low magnification bright field (BF) (a) and medium angle annular dark field (MAADF) (b) images revealing the $a/c$ phase of the \pto\ layer, with the $a/c$ domain walls inclined in the $(011)_{\mathrm{pc}}$ or $(0\overline{1}1)_{\mathrm{pc}}$ planes. (c-d) High resolution images around two $a$-domains with different domain wall orientations, showing $a_1$ and $c_1$ domains separated by domain walls in the $(011)_{\mathrm{pc}}$ planes (c), and $a_2$ and $c_2$ domains separated by domain walls in the $(0\overline{1}1)_{\mathrm{pc}}$ planes (d). The arrows indicate the Ti displacements. (e) Schematic representation of the domain pairs. (f) FFT pattern of the \pto\ layer where the butterfly shape of the (001)$_{\mathrm{pc}}$ is highlighted in (g). Squares in (g) represent the regions of interest used for the image reconstruction, the top squares corresponding to signal from $a$-domains, and the bottom ones from the $c$-domains. (h-i) Images reconstructed from the intensity of the FFT filtered pattern for the $a$-domains (h) and $c$-domains (i). (j-k) FFTs for areas corresponding to $a_1$ and $a_2$ domains are overlapped in (j), $c_1$ and $c_2$ in (k).} 
\end{figure}

A more direct way to image the domain structure is the use of cross-sectional scanning electron microscopy (STEM) (Figure~\ref{fig:Figure4}). Atomically resolved high angle annular dark field (HAADF)-STEM imaging allows us to observe the lattice and atomic displacements at the atomic level. The samples were cut and prepared for the STEM measurement so as to obtain slices in the plane defined by the [010]$_{\mathrm{pc}}$ (horizontal direction) and [001]$_{\mathrm{pc}}$ (vertical direction) axes of \dso , imaging along the [100]$_{\mathrm{pc}}$ zone-axis. In the 90 u.c. thick \pto\ in Figure~\ref{fig:Figure4}, the $a$-domains are clearly visible as very narrow domains separated from the larger $c$-domains by domain walls in the $(011)_{\mathrm{pc}}$ and $(0\overline{1}1)_{\mathrm{pc}}$ planes. 

Focusing on a region with the domain walls in the $(011)_{\mathrm{pc}}$ planes (Figure ~\ref{fig:Figure4}(c)), we observe the tilted $a$ and $c$-domains, corresponding to an $a_1$-domain surrounded by $c_1$-domains. In the case of domain walls in the $(0\overline{1}1)_{\mathrm{pc}}$ planes (Figure ~\ref{fig:Figure4}(d)), we observe an $a_2$-domain surrounded by $c_2$-domains. These correspond to the $a_1/c_1$ and $a_2/c_2$ pairing observed in the nanodiffraction data (Figure~\ref{fig:Figure3}). 

Interestingly, in these two different regions, the polarisation in the $c$-domains is always oriented down (in the [001]$_{\mathrm{pc}}$ direction), while the polarisation in the $a$-domains is always oriented to the right (in the [010]$_{\mathrm{pc}}$ direction). %A possible scenario is proposed in Supplementary Materials Section~\ref{Section:SI_TEM_ac}.

At a larger scale, the periodic pattern is directly visible in real space and the period can be estimated by taking the FFT of the image, as shown in Figure~\ref{fig:Figure4}(f). Looking at the (001)$_{\mathrm{pc}}$ Bragg spot in the obtained reciprocal space map (Figure~\ref{fig:Figure4}(g)), one can see the superstructure with a periodicity corresponding to 40-45 nm, in perfect agreement with the values obtained from the RSM (see Figure~\ref{fig:Figure5} for the comparison between values obtained by the different techniques).

In Figure~\ref{fig:Figure4}(g), one also recognizes the butterfly shape as observed by XRD and nanodiffraction. By selecting different regions of interest, it is then possible to reconstruct the direct space images in Figure~\ref{fig:Figure4}(h-i) revealing the mapping of the $a$ and $c$-domains ($a_1$ and $c_1$ in pink, $a_2$ and $c_2$ in green). Although the reconstructed image is sharper for the $a$-domains (Figure~\ref{fig:Figure4}(h)) than for the $c$-domains (Figure~\ref{fig:Figure4}(i)), it allows us to confirm at a local scale and with direct imaging the result obtained by nanodiffraction: the $a$ and $c$-domains organize themselves, not only in pairs, but in larger regions composed of several $a_1$/$c_1$ pairs alternating with larger regions composed of several $a_2$/$c_2$ pairs, forming superdomain structures.

\section{Discussion}
\label{Section:Discussion}

\begin{figure}[!htb]
\includegraphics[width=\linewidth]{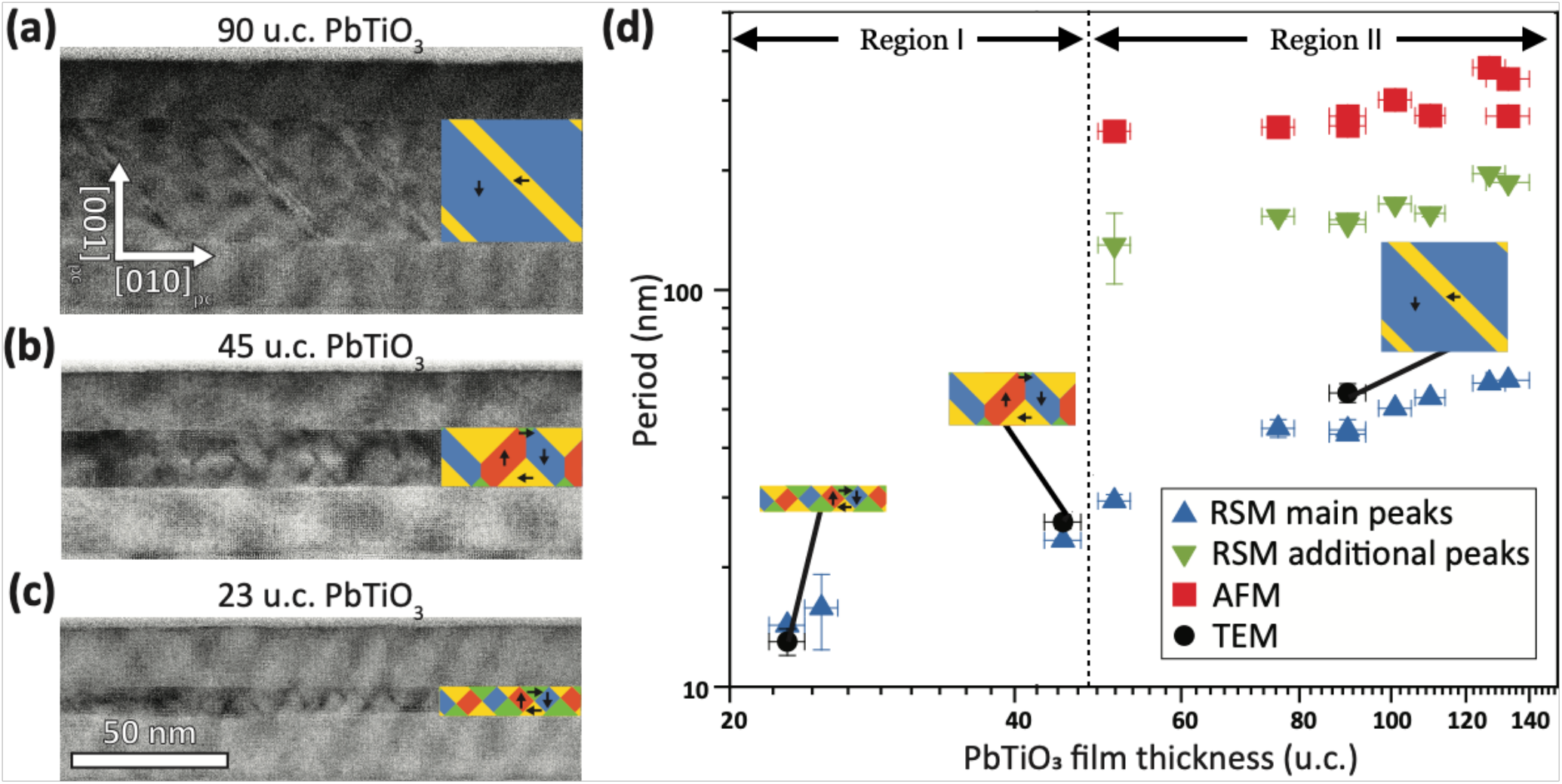}
\caption{\label{fig:Figure5} Summary of the domain evolution along [010]$_{\mathrm{pc}}$ as a function of \pto\ thickness. (a-c) STEM-BF images revealing the domain walls and domain pattern evolution from flux-closure like to $a/c$ as the \pto\ film thickness increases. (d) Evolution of the periods obtained from the AFM topography images (red dots), via RSM (blue and green triangles for the main period and additional period respectively), and from STEM images (purple diamonds). From this Figure, we see a good match between the values obtained by STEM and RSM. We also see that two different regions can be defined: one region where no pattern is visible in the topography and no additional peaks appear in the RSM for the thinner \pto\ films (45 u.c. and lower), and one region where the topography displays tilts and trenches, and where additional peaks appear in the RSM for the thicker \pto\ films (50 u.c. and higher).} 
\end{figure}

The different periods observed in this work along [010]$_{\mathrm{pc}}$ are summarised in Fig.~\ref{fig:Figure5}(d): from the AFM topography, the RSM, and the STEM images. From this Figure, we see a good match between the values obtained by STEM and RSM. We also see that two different regions can be defined: region I where no pattern is visible in the topography for the thinner \pto\ films (45 u.c. and lower), and region II where the topography displays tilts and trenches, and where additional peaks appear in the RSM for the thicker \pto\ films (50 u.c. and higher). 

The STEM measurements show that the sample in region II has a clear $a/c$-phase domain structure (Fig.~\ref{fig:Figure5}(a) - 90$\pm$4 u.c. thick \pto ), while the two samples measured in region I show a more complex domain pattern (Fig.~\ref{fig:Figure5}(b) - 45$\pm$2 u.c. and (c) - 23$\pm$1 u.c. thick \pto\ layers). In these images, domain walls are clearly visible in the \pto\ layers, forming an $a/c$ pattern for the thicker \pto\ layer and transforming into a flux-closure like pattern for the thinner \pto\ layers, sharing similarities to what was observed in \pto /\sto\ heterostructures on \gso ~\cite{Tang-Science-2015}.

By performing FFT filtering for these three films, as was done for the 90 u.c. thick \pto\ sample in Figure~\ref{fig:Figure4}, we extracted the periodicity of these different domain configurations and found 40-45 nm for the 90 u.c. thick \pto , 25-30 nm for the 45 u.c. thick \pto , and 16 nm for the 23 u.c. thick \pto , in perfect agreement with the values obtained from the RSM (see Figure~\ref{fig:Figure5}(d) - black dots). 

Nanodiffraction measurements allowed us to spatially resolve the $a/c$ pattern in the sample with 90 u.c. thick \pto\ and demonstrated the coupling of the tilts of the $a$- and $c$-domains, as also observed by STEM. 

Nanodiffraction also demonstrated the organisation of these domains at a larger scale. From the STEM measurements combined with the spatially resolved XRD measurements, we conclude that in the $a/c$-phase, the $a/c$-domains organise in ``superdomains'', i.e. regions composed of either $a_1/c_1$, $a_2/c_2$, $a_3/c_3$, or $a_4/c_4$ domains. This gives rise to an additional periodicity that can be detected as additional peaks in the RSM. From the PFM measurements in a sample without the top \sro\ electrode, we relate these additional periodic peaks to the 180$^\circ$ ferroelectric domains ($c^+/c^-$). Although we could not confirm it directly, one possibility is that these $c^+/c^-$ domains correspond to the superdomains observed with different $a/c$ tilts.

Looking at the evolution of the periods as a function of film thickness, we observe that the period $p$ of the ferroelastic domain structure in the \pto\ layers decreases with decreasing film thickness $t$ as $p\sim t^n$ with $n=0.8\pm0.1$. This is very close to the value obtained by Nesterov~\etal~\cite{Nesterov-APL-2013} ($n=0.68 \pm 0.3$). It also compares well with the results obtained in different works (see Supporting Information, Section~\ref{Section:SI_periods}).

When considering the energetics at play in ferroelastic domains in epitaxial ferroelectric and ferroelastic films, Pertsev~\cite{Pertsev-JAP-1995} showed that the dependence of the equilibrium domain period on film thickness varies from linear to the usual square root law at very large film thickness, as the domain wall energy density varies with film thickness. At very low thicknesses, the domain size can even increase as the film thickness is reduced~\cite{Pertsev-JAP-1995,Kopal-Ferroelectrics-1997,Huang-JAP-2013}. Although we did not observe such a turnover of domain size for our samples with lower film thickness, we believe this mechanism is at play in our materials and could explain the existence of the two different regions. In region II, the samples are in the $a/c$-phase with the period expected for this epitaxial strain. They form superdomains, probably controlled by the $c^+/c^-$ domain configuration determined by the screening of the depolarisation field and electrostatic boundary conditions. Both of these patterns in region II have a period that decreases with film thickness, following the Kittel law for the superdomains, and following an intermediate behavior between Kittel's square-root law and Pertsev's linear law for the ferroelastic $a/c$-domains. In this region, the change between $c^+$ and $c^-$ can easily take place on each side of the $a$-domains, without additional cost of formation of 180$^\circ$ domain walls. However, as the film thickness is reduced further, we enter in region I into a regime where the $a/c$-domain size should increase exponentially as the film thickness is reduced, going towards a monodomain state. However, the cost of the depolarisation field is still increasing and needs to be compensated by $c^+/c^-$ domains. In this regime, many 180$^\circ$ domain walls would then have to be created, at a large energy cost. We observe that instead of this configuration, the domains end up in a more complex polarisation texture, similar to flux-closure pattern, with a period following the behavior of the $a/c$-domain pattern at larger thickness, and without any additional peaks or superdomain structure. This new complex polarisation pattern appearing at lower thickness is thus the best compromise between the behavior expected for ferroelectric vs. ferroelastic domains, reducing the cost of strain and depolarisation field at the same time. This explains why in this regime, it is possible to observe very complex patterns such as flux-closure~\cite{Tang-Science-2015,Hadjimichael-NatMat-2021}, vortices~\cite{Yadav2016,Shafer2018,Damodaran2017}, supercrystals~\cite{Hadjimichael-NatMat-2021} and incommensurate spin (polarisation) structures~\cite{Rusu-Nature-2022}, all with the period determined by the strain from the substrate and the layer thickness.

\section{Conclusion}

Our work demonstrates the presence of two regimes with different ferroelectric/ferroelastic domain configurations. For the smaller \pto\ thicknesses, the combination of the effect of strain and electrostatic boundary conditions gives rise to a complex domain configuration where flux-closure, vortices or supercrystal configurations can develop. For larger \pto\ thicknesses, the $a/c$-phase induced by epitaxial strain is recovered, with large superdomains appearing to screen the depolarisation field. The large structural distortions associated with the ferroelastic domains propagate through the top \sro\ layer, creating a modulated structure that extends beyond the ferroelectric layer thickness, allowing domain engineering in the top \sro\ electrode~\cite{Lichtensteiger-SRO-ToBeSubmitted-2023}. These domain structures not only change the properties of the ferroelectric itself, but can also be used to change the properties of other materials through electrostatic and structural coupling. 

\section{Experimental techniques}
\label{ExperimentalTechniques}

\subsection{Sample growth}
\label{section:Growth}
All the samples were deposited using our in-house constructed off-axis radio-frequency magnetron sputtering system, equipped with three different guns allowing the deposition of heterostructures and solid-solutions of high crystalline quality, composed of up to three different materials: \pto , \sto\ and \sro .
\pto\ thin films are typically deposited around 560$^\circ$C and 580$^\circ$C, in 180 mTorr of a 20:29 O$_2$/Ar mixture, at a power of 60 W, and using a Pb$_{1.1}$TiO$_3$ target with 10\% excess of Pb to compensate for its volatility.
\sro\ electrodes were deposited {\it in situ} from a stoichiometric target at 640$^\circ$C in 100 mTorr of O$_2$/Ar mixture of ratio 3:60, at a power of 80 W. 
Huettinger PFG 300 RF power supplies are used in power control mode. The sample holder is grounded during deposition, but the sample surface is left floating. 

\subsection{Atomic force microscopy}

Topography measurements were performed using a {\it Digital Instrument} Nanoscope Multimode DI4 with a {\it Nanonis} controller. Piezoresponse force microscopy measurements under ambient conditions to image the intrinsic domain patterns were performed on an {\it Asylum Research Cypher} or {\it MFP-3D} atomic force microscopes.

\subsection{X-ray diffraction}

In-house XRD measurements were performed using a {\it Panalytical X'Pert} diffractometer with Cu K$\alpha_1$ radiation (1.5405980 \AA ) equipped with a 2-bounce Ge(220) monochromator and a triple axis detector in our laboratory in Geneva. The $\theta$-2$\theta$ scans were analysed using the {\it InteractiveXRDFit} software~\cite{Lichtensteiger-JApllCryst-2018}. This XRD system is also equipped with a PIXcel1D detector, used for faster acquisition of the reciprocal space maps. 

The 90 u.c. thick \pto\ layer sample was further analysed using scanning x-ray diffraction microscopy (SXDM)~\cite{Hadjimichael-PRL-2018} on the ID01 beamline at the European Synchrotron Radiation Facility (ESRF)~\cite{Leake2019}. An incident x-ray energy of 9.5 keV was selected using a Si(111) double crystal monochromator with resolution $\Delta\lambda/\lambda = 10^{-4}$. A $0.7$ $\mu$m-thick tungsten Fresnel zone plate with 300 $\mu$m diameter and 20 nm outer zone width was used to focus the x-ray beam down to a spot size of approx. $30 \times 30$ $\mathrm{nm}^2$ FWHM, as measured via a ptychography scan of a known reference object~\cite{Leake2019}. The sample was placed at the $002$ diffraction condition and raster scanned relative to the focused beam in 25 nm steps over a $4 \times 4$ $\mu\mathrm{m}^2$ area using a commercial piezo scanner. A two-dimensional MAXIPIX detector positioned approx. 0.4 m downstream of the sample stage was exposed for 12 ms at each sample position. The procedure was repeated at different incidence angles about the $002$ condition to obtain a 5-dimensional dataset describing the diffracted intensity as a function three reciprocal and two direct space dimensions, i.e. $I(Q_{[100]_{\mathrm{pc}}},Q_{[010]_{\mathrm{pc}}},Q_{[001]_{\mathrm{pc}}},R_{[100]_{\mathrm{pc}}},R_{[010]_{\mathrm{pc}}})$. We note that reaching the $002$ diffraction condition entails forming a $\sim$19$^\circ$ angle with the (00L) planes; since measurements were performed with x-ray beam parallel to $[100]_{\mathrm{pc}}$, the direct space resolution is degraded along this direction, and features appear elongated parallel to it.

\subsection{Scanning transmission electron microscopy}

Cross-sectional lamella prepared by focus ion beam  allow the imaging of domain structures by scanning transmission electron microscopy (STEM). Experiments were acquired on Nion Cs-corrected UltraSTEM200 at 100 kV operating voltage. A convergence angle of 30 mrad was used to allow high-resolution atomic imaging with a typical spatial resolution of 1 \AA. Three imaging detectors in the STEM are used to simultaneously obtain bright field (BF), annular bright field (ABF) or medium angle annular dark field (MAADF), and high angle annular dark field (HAADF) images.

For ABF-MAADF imaging, the inner-outer angles can be continuously adjusted between 10-20 to 60-120 mrad. Most ABF images were collected with 15-30 mrad and MAADF images with 40-80 mrad angular ranges. 

We determine the periodicity of the superstructures in the \pto\ layer by measuring the distances between the additional reciprocal space spots obtained after FFT. The accuracy of the measurement was estimated by considering the diffraction spot extension as the lower and upper limit for the superstructure length estimation.

\section{Data availability}

The data that support the findings of this study are available at Yareta (DOI).

\section{Acknowlegements}

The authors thank Kumara Cordero-Edwards and Christian Weymann for support and discussions.

This work was supported by Division II of the Swiss National Science Foundation under project 200021\_200636. STEM experiments were supported by the EU Horizon research and innovation program under grant agreement ID 823717-ESTEEM3. M.H. acknowledges funding from the SNSF Scientific Exchanges Scheme (Grant Number IZSEZ0\_212990).

\section{Author contributions}

C.L, M.H., P.P., A.G. and J.M.T. designed the experiment. C.L., M.H. and L.T. grew the samples and conducted the AFM and XRD measurements and analysis. E.Z. conducted the synchrotron x-ray nanodiffraction measurements and analysis. C.-P.S. and A.G. conducted the STEM measurements and analysis. I.G. performed additional STEM analysis. C.L., M.H., I.G. and P.P. wrote the manuscript with contributions from all authors. All authors discussed the experimental results and models, commented on the manuscript, and agreed on its final version.

%\newpage
\section{Bibliography}
%\sloppy
%\normalem
%\printbibliography
\bibliography{biblio}
\bibliographystyle{naturemag}

%\newpage
\section*{Supplementary Materials}

\renewcommand{\thepage}{S\arabic{page}} 
\setcounter{page}{1}
\renewcommand{\thesection}{S\arabic{section}}  
\setcounter{section}{0}
\renewcommand{\thetable}{S\arabic{table}} 
\setcounter{table}{0}
\renewcommand{\thefigure}{S\arabic{figure}}
\setcounter{figure}{0}

\section{Observation of $c^+/c^-$ superdomains in a sample without the top \sro\ electrode}
\label{SI:topo_PFM_XRD_uncapped}

\begin{figure}[!htb]
\includegraphics[width=\linewidth]{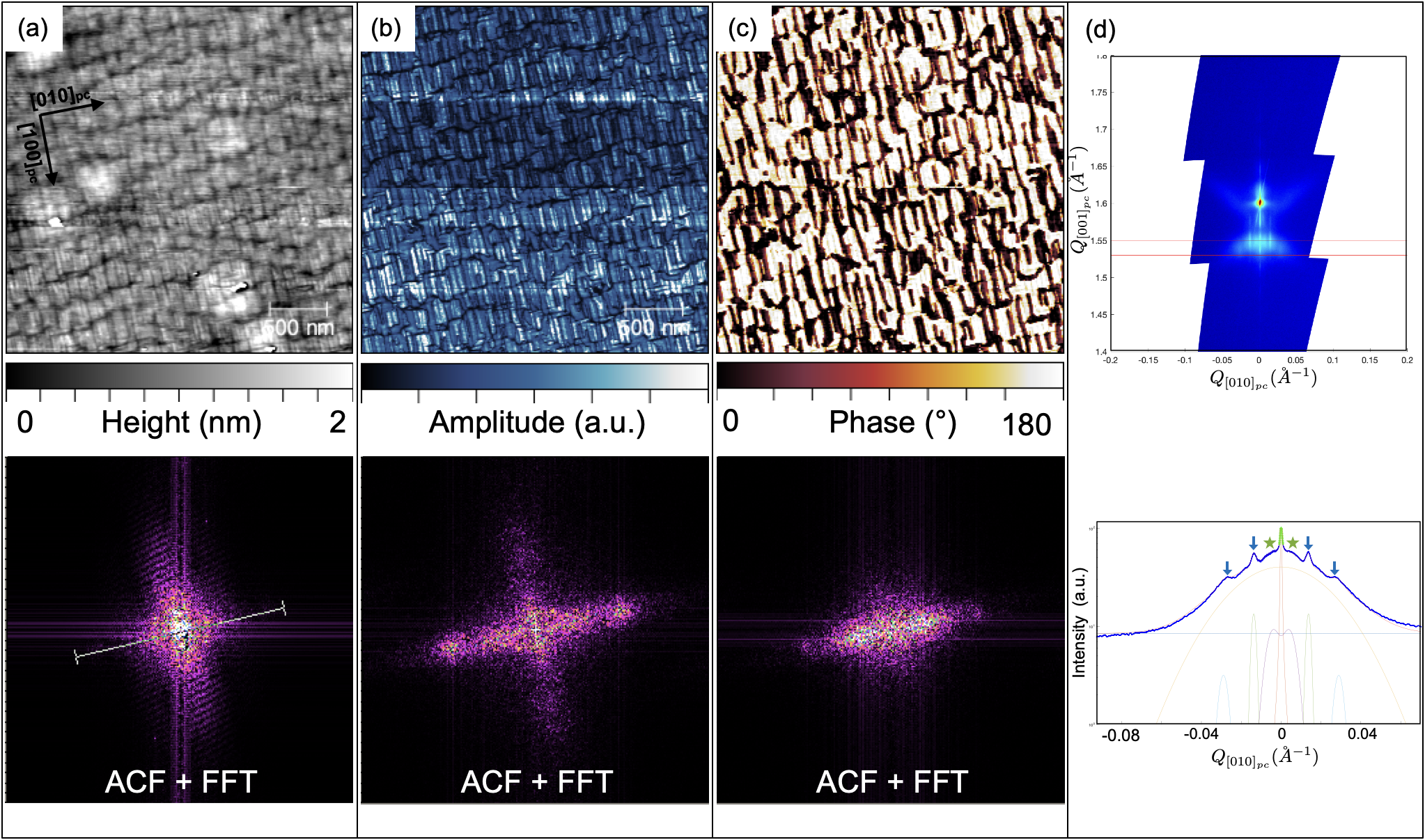}
\caption{\label{fig:SI_topo_PFM_XRD_uncapped} Analysis of a sample without top \sro\ electrode - 90 $\pm$ 4 u.c. \pto\ on 53 $\pm$ 3 u.c. \sro\ on \dso\ substrate. (a) Topography image and ACF+FFT analysis. (b) PFM vertical Amplitude image and ACF+FFT analysis. (c) PFM vertical phase image and ACF+FFT analysis. (d) RSM and cut. See text for more details.}
\end{figure}

To allow for direct visualisation of domain configuration using piezo-response force microscopy (PFM), we also grew a sample without the top \sro\ electrode: 90 $\pm$ 4 u.c. \pto\ on 53 $\pm$ 3 u.c. \sro\ on \dso\ substrate. In Figure~\ref{fig:SI_topo_PFM_XRD_uncapped}, we show the topography image (a), where we can observe the trenches at the top of the \pto\ layer, very similar to what we observe in the sample of the same \pto\ thickness with the top \sro\ electrode. The anisotropic pattern has a period of $\sim$266 nm along [100]$_{\mathrm{pc}}$ and $\sim$46 nm along [010]$_{\mathrm{pc}}$ (ACF+FFT analysis of the topography image). In (b) we show the PFM vertical amplitude, where we see dark lines corresponding to 180$^\circ$ domain walls and/or $a$-domains. ACF+FFT analysis of the amplitude image gives a period of $\sim$280 nm along [100]$_{\mathrm{pc}}$ and $\sim$48 nm along [010]$_{\mathrm{pc}}$, following the values obtained on the topography image. In (c) we show the PFM vertical phase, with a clear signature of superdomains where the $a/c$ domains organise in regions where the out-of-plane polarisation points up, alternating with regions where the out-of-plane polarisation points down. From the ACF+FFT analysis of this PFM vertical phase image, we extract a period of 1175 $\pm$ 150 nm along [100]$_{\mathrm{pc}}$ and 145 $\pm$ 6 nm along [010]$_{\mathrm{pc}}$ for the $c^+/c^-$ superdomains. In (d), we show the RSM obtained around (001)$_{\mathrm{pc}}$ in the $Q_{[001]_{\mathrm{pc}}}-Q_{[010]_{\mathrm{pc}}}$ plane, showing the typical butterfly shape of the $a/c$-phase and the additional peaks of the periodic pattern. From the cut at $Q=1.54$ \AA$^{-1}$ ($c$-domains), we extract the intensity profile showing the periodic peaks indicating a periodicity of 45 $\pm$ 2 nm. This value matches perfectly with the values obtained from the analysis of the topography and amplitude images. From the intensity cut in the RSM, we also see the shoulders (indicated by green stars in Figure~\ref{fig:SI_topo_PFM_XRD_uncapped}(d)), corresponding to an additional period of 150 $\pm$ 2 nm. This value corresponds to that obtained from the analysis of the PFM phase image, confirming that the additional peaks come from the arrangement of the out-of-plane polarisation in $c^+/c^-$ superdomains.

\section{Deviation from Kittel law: Comparison with ferroelastic and flux-closure domains
in other \pto\ thin films and superlattices on \dso }
\label{Section:SI_periods}

\begin{figure}[!htb]
\includegraphics[width=0.95\linewidth]{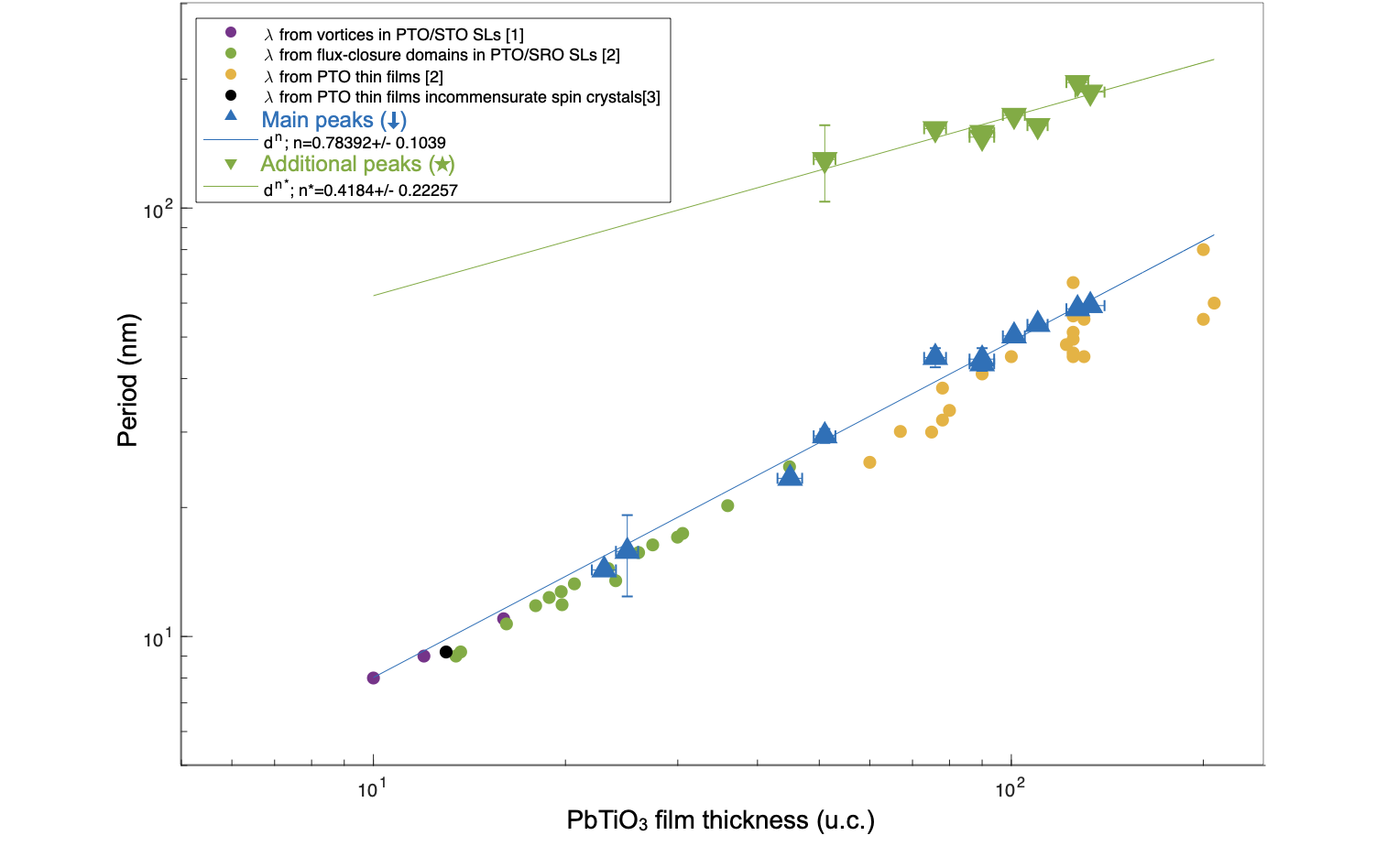}

\caption{\label{fig:SI_periods} Domain period versus film thickness: deviation from Kittel law. Comparison with ferroelastic and flux-closure domains in other \pto\ thin films and superlattices on \dso. Purple: vortices periods in \pto /\sto\ superlattices from Yadav \etal ~\cite{Yadav2016}, Schafer \etal ~\cite{Shafer2018} and Damodaran \etal ~\cite{Damodaran2017}. Green: flux-closure domains in \pto /\sro\ superlattices from Hadjimichael \etal ~\cite{Hadjimichael-NatMat-2021}. Yellow: domain patterns in \pto\ thin films from Hadjimichael \etal ~\cite{Hadjimichael-NatMat-2021}. Black: domain period from \pto\ thin films incommensurate spin crystals from Rusu \etal ~\cite{Rusu-Nature-2022}.}
\end{figure}

Comparing to other periods found in the literature to our results (Figure~\ref{fig:SI_periods}), we see that the trend we observe is universally followed: the period in the \pto\ layers is the same in all the different heterostructures and is fixed by the epitaxial strain imposed by the \dso\ substrate, whether in thin films or superlattices, forming $a/c$-domains or more complex polarisation textures like flux-closure, vortices, supercrystals or incommensurate spin crystals.

\end{document}